\documentclass[sigconf]{acmart}
\AtBeginDocument{%
  \providecommand\BibTeX{{%
    \normalfont B\kern-0.5em{\scshape i\kern-0.25em b}\kern-0.8em\TeX}}}

\copyrightyear{2022}
\acmYear{2022}
\setcopyright{rightsretained}\acmConference[CHI '22]{CHI Conference on Human Factors in Computing Systems}{April 29-May 5, 2022}{New Orleans, LA, USA}
\acmBooktitle{CHI Conference on Human Factors in Computing Systems (CHI '22), April 29-May 5, 2022, New Orleans, LA, USA}
\acmConference[CHI 2022]{ACM CHI Conference on Human Factors in Computing Systems}{April 30--May 06}{New Orleans, Louisiana, USA.}
\acmPrice{}
\acmDOI{}
\acmISBN{}

\usepackage{graphicx}
\usepackage[utf8]{inputenc}
\usepackage{amsmath}
\usepackage{units}
\usepackage{xcolor}
\usepackage{booktabs}
\usepackage{textcomp}
\usepackage{multirow}

\usepackage{microtype}        
\usepackage{ccicons}          

\author{Kashyap Todi}
\email{kashyap.todi@gmail.com}
\affiliation{%
\institution{Meta Reality Labs}
\country{USA}
}
\author{Ben Lafreniere}
\email{benlafreniere@fb.com}
\affiliation{%
\institution{Meta Reality Labs}
\country{Canada}
}
\author{Tanya Jonker}
\email{tanya.jonker@fb.com}
\affiliation{%
\institution{Meta Reality Labs}
\country{USA}
}

\begin{document}

\title{Computational Adaptation of Extended Reality Interfaces Through Interaction Simulation}



\begin{abstract}

Adaptive and intelligent user interfaces have been proposed as a critical component of a successful extended reality (XR) system \cite{chi20jonker}.
In particular, a predictive system can make inferences about a user and provide them with task-relevant recommendations or adaptations. 
However, we believe such adaptive interfaces should carefully consider the overall \emph{cost} of interactions to better address uncertainty of predictions.
In this position paper, we discuss a computational approach to adapt XR interfaces, with the goal of improving user experience and performance.
Our novel model, applied to menu selection tasks, simulates user interactions by considering both cognitive and motor costs.
In contrast to greedy algorithms that adapt based on predictions alone, our model holistically accounts for costs and benefits of adaptations towards adapting the interface and providing optimal recommendations to the user.
\end{abstract}




\maketitle

\section{Introduction}

Extended reality (XR) is a growing area of post-desktop computing, spanning virtual reality (VR), augmented reality (AR), and mixed reality (MR).
In VR, users are immersed in complex 360\textdegree{} environments where digital content can be freely accessed.
In AR and MR, users can seamlessly interact with both the real world and virtual content, blurring the boundaries between them.
This is radically different from current interaction paradigms where digital interfaces occupy their own dedicated physical spaces (screens) and offer users optimized and performant input methods to interact with them.
It is still unclear what new interaction capabilities and interfaces can best support the user and enable them to better interact with content here.
Given recent advancements in computational interaction \cite{oulasvirta2018computational,li2021artificial}, we see great potential for applying or developing computational approaches to generate or adapt XR user interfaces (UIs).
In contrast to static UIs, these approaches could dynamically adapt the UI to a user's context and activity, thereby improving interactions.

In this position paper, we introduce a model that can simulate a user's interactions with an XR interface, with the goal of selecting optimal UI adaptations.
We assume that the system has a predictive model that provides a probability distribution over all possible actions in the interface.
In a menu interface, for instance, the system could predict the selection probability for each of the menu items. 
Prior work has developed methods and models for predicting users' intentions given some prior information such as the user's interaction history and their context (e.g. \cite{10.1145/3172944.3172949, 10.1145/3448018.3458008,10.1145/3332165.3347933,10.1145/2037373.2037467}).
Given a probability distribution over available actions, a predictive system can then generate the UI.
For example, it could generate a simplified menu that prioritizes items with high probability (e.g. \cite{10.1145/3308557.3308705}), preferentially reveal items \cite{10.1145/1518701.1518951}, gradually reorganize items \cite{10.1145/3411764.3445497}, or augment the standard menu interface with recommendations (e.g. \cite{sears1994split}).

A key aspect of our proposed model is that it considers various \emph{costs} during interaction. 
Recent work has studied the cognitive costs and recovery time associated with recognition errors \cite{10.1145/3472749.3474735}, and planning-based approaches have been developed to estimate long-term costs and benefits of adaptations  \cite{10.1145/3411764.3445497}. 
Here, we study cognitive and interaction costs in the context of a predictive system that makes adaptive recommendations based on contextual predictions about the user's intents\footnote{In this paper, we assume that contextual predictions are available to the adaptive system; we do not develop this predictive model here.}.
We argue that a `greedy' recommender approach -- one that exclusively prioritizes high probability actions -- might be sub-optimal when used directly because it does not take into account the interface design and potential adverse interaction costs of inaccurate predictions. 
We develop a simulation model of a user's interactions with a hierarchical menu.
The model takes into account both \emph{benefits} of a suitable adaptation and \emph{costs} of inaccurate predictions.
We consider three key time-based costs of interactions -- inspection cost, selection cost, and correction cost -- to compute the \emph{utility} of any given adaptation.
In contrast to a greedy baseline that always selects the highest-probability action, our model attempts to minimize the overall selection time.
The model is instantiated in a menu-based XR system, enabling it to adaptively select a utility-maximizing starting point in the menu hierarchy.
We demonstrate the model's predictions in a system with illustrative scenarios.

\section {Overview}

\subsection{Background: Adaptive User Interfaces}

In this paper, we discuss XR systems that can adapt their interface to a user.
By observing the user's context, such systems can make probabilistic predictions about their future intentions \cite{chi20jonker}.
An \emph{adaptation policy} determines how the interface should adapt given a contextual probability distribution.
Prior works have explored several policies, such as frequency, recency, and access rank, among others (e.g. \cite{10.1145/3308557.3308705,mitchell1989dynamic, gajos2006exploring}). 
However, these typically assume that the system's predictions are accurate without fully considering the cost of inaccurate predictions.
As such, they can be `greedy' in that they aim to optimize adaptations solely based on these predictions by recommending highest probability actions.
When making adaptations, they do not fully take into account the UI design or the fact that the user might have learned parts of it.
To address these gaps, in the present work, we introduce a model-based approach for adapting XR interfaces.
The system considers the utility of adaptations given contextual observations of the user to select an optimal adaptation that can both maximize the benefits of an accurate prediction and minimize the costs incurred otherwise.

\subsection{General Approach: Model-Based Simulation}
In our model-based approach, the goal is to estimate the \emph{utility} \cite{savage1972foundations,10.1145/302979.303030} of an adaptation given a contextual probability distribution over actions supported by the interface.
This utility is a combination of the \emph{benefits} provided by an adaptation and the \emph{costs} incurred due to inaccurate predictions.
As such, it is tightly coupled to both the system's predictions and the user interface design itself.
The key to computing utility is simulation of the user's interaction sequence when completing a given task.
We consider target acquisition tasks, where a user interacts with the UI to find and select a \emph{target item}, which is common when interacting with typical graphical UIs such as menus, toolbars, and interactive applications.
Each interaction sequence -- navigating from a starting point to the target item -- has a \emph{total cost} associated with it. 
This could be quantified by, for example, the time taken to complete the interaction sequence, cognitive load, or physical exertion.
Conversely, an adaptation has a \emph{benefit} associated with it; this is the cost that is circumvented through adaptation of the starting point.

For any general adaptive interface, the utility of an adaptation is a combination of the benefit provided by the adaptation and the costs incurred due to inaccurate adaptations where the user's target item differs from the one favored by the adaptation.
By modeling the sequence of actions taken by a user to complete an interaction task and estimating the costs incurred, an adaptive system can effectively use simulations to estimate the utility of each possible adaptation.
It can then select the utility-maximizing adaptation to optimize the interface given the user's current context. 
We develop one such adaptive system and elaborate upon the simulation model that drives interface adaptations here.

\section {Model Formulation}\label{sec:model}


\subsection{Application: Adaptive XR System}\label{sec:system_overview}

Menus have received extensive attention in HCI research as they are widely used and adaptation has potential to improve usability \cite{10.1145/2501988.2502024}.
Our simulation-based interaction model studies the case of \emph{hierarchical menus}.
In our design, the user can traverse the menu hierarchy, where items are grouped semantically into sub-menus, and select items using \emph{selection} actions.
A dedicated \emph{correction} action (e.g. back button or gesture) enables the user to undo a selection, or go one step up the hierarchy to a parent menu.
While static menus always provide the top-level item (root node) as the starting point for interaction, we develop an \emph{adaptive menu} that can dynamically select any item as a starting point.
This could either be the root node, any intermediate sub-menu, or a leaf item (final action).
\autoref{fig:menu_example} illustrates the adaptive system using an example menu.
Our simulation model adapts such a menu with the objective of minimizing \emph{selection time}.

\subsection{Interaction Cost Parameters}

Previous literature has developed models to explain a user's search process within linear menus \cite{10.1145/3411764.3445497}.
We build upon prior work to formulate a model for simulating user interactions with our hierarchical menu-based XR interface.
The model takes into account cognitive and motor components of the interaction, given by the three \emph{time-based cost} parameters, to estimate the total cost of a specific interaction event.

\subsubsection{Inspection Cost ($T_{inspect}$):}
When traversing a menu, the user needs to first visually inspect the interface to find relevant items.
Each visual inspection has a cost associated with it.
In this work, we consider that inspection cost is item-independent, and use a constant time-based cost of inspection $T_{inspect}$.



\subsubsection{Selection Cost ($T_{select}$):} This parameter defines the cost associated with taking a selection action.
As the user inspects the interface and finds relevant sub-menus or the target item, they provide input by selecting the item.
This selection could be using a point-and-click technique, a dedicated gesture, or other input modalities.
The cost associated with this action is directly dependent on the input technique, and is influenced by factors such as input precision, encumbrance, and noise.

\subsubsection{Correction Cost ($T_{correct}$):}
During interaction, the user might need to take corrective actions either when they make an incorrect selection or when the system presents an adaptation that is not suitable towards the user's actual intention.
Such corrective actions also have costs associated with them.
Similar to the selection cost, the input technique used to enable corrective actions influences the correction cost.

\subsection{Simulating User Interactions}

We formulate a model for simulating user interactions with the described adaptive XR interface.
Given a starting menu item as stimulus, it produces the set of actions required to acquire a target item.
We consider that a user adopts a serial search strategy, where they inspect each menu item until they find their target item or they reach the end of the menu.

\subsubsection{Search-and-Select:} During a target acquisition task, the user inspects (reads) the menu interface until they either find a sub-menu item that contains their target item or the target item itself; upon finding this item, they make a selection.
The total time cost for this sequence of actions ($T_{search}(i_l)$) is given by:
\begin{equation}
    T_{search}(i_l) = \sum_{j=1}^{l} T_{inspect} + T_{select}
\end{equation}
where $i_l$ is either the target sub-menu or target item and $l$ is the location of $i_l$.
To recursively search and select within sub-menus, the above search process is repeated and costs are accumulated.

\begin{figure*}[t]
    \centering
    \includegraphics[width=.95\textwidth]{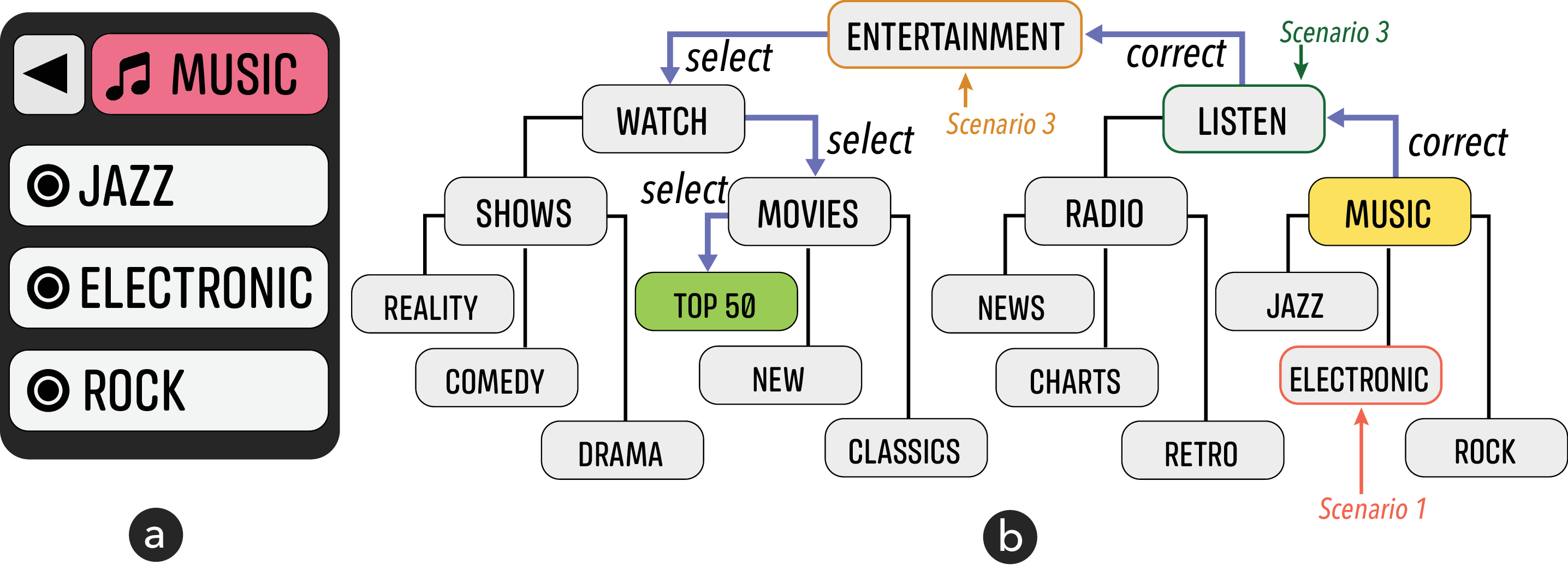}
    \caption{(a) An example application UI, with `Music' as the currently selected item. (b) Hierarchical structure of the application menu, with `Entertainment' as the top-level item (root). The structure shows the path from the current item (`Music', highlighted in yellow) to the target item (`Top 50', highlighted in green). To complete this target acquisition task, the user needs to make two corrections followed by three selections. The tree diagram is also annotated to show the selected adaptations for the three scenarios.}
    \label{fig:menu_example}
    \Description{A two panel figure. Left: An illustration of a UI with a title, back button, and list with 3 items. Right: A tree diagram of a hierarchical menu.}
\end{figure*}

\subsubsection{Backtracking:} If the user encounters a sub-menu that does not contain the target item, they need to take a sequence of corrective actions, called \emph{backtracking}, to traverse the menu hierarchy and find the appropriate parent menu before beginning the above \emph{search-and-select} process.
Each corrective action incurs a \emph{correction cost}, following which, items in the parent sub-menu are displayed.
The user now serially searches for a relevant item within this sub-menu; if they fail to find such an item, they repeat these backtracking steps until they encounter a relevant item.
The total backtracking cost to recursively find an item related to the target item is given by:
\begin{equation}
    T_{backtrack} = \sum_{k=1}^{n} (T_{correct} + (l_k \cdot T_{inspect}))
\end{equation}
where $n$ is the number of sub-menus and $l_k$ is either the position of the relevant item within a sub-menu $k$ or the number of items in sub-menu $k$ if the relevant item is not present.

\subsubsection{Total Interaction Cost:}
Using the two components above, search-and-select and backtracking, we can compute the total cost of any target acquisition task.
When the user is presented with an initial menu item $k$ and has a target $l$, this cost is given by:
\begin{equation}
    T(k,l) = T_{backtrack} + T_{search}
\end{equation}
where $T_{backtrack}$ is the total time cost required to backtrack to the closest common parent item of $k$ and $l$, and $T_{search}$ is the time cost for inspecting the menu and making selections until the target item $l$ is acquired.

\subsection{Selecting Adaptations}

With the above formulation, we can predict the total cost for any interaction sequence.
As described in \autoref{sec:system_overview}, the system adapts the interface by selecting a suitable starting point $k$, given a probability distribution $p$ over all leaf menu items.
Each adaptation has a \emph{utility} associated with it, which is the combination of the \emph{benefit} and \emph{costs} of an adaptation $k$:
\begin{equation}
Utility_{k} = \left(\sum_{i=0}^n p_i \cdot T(k,i)\right) -   (p_{k} \cdot Benefit(k))
\end{equation}
where $n$ is the number of leaf items, $T(k,i)$ is the cost for starting from item $k$ when the target is $i$, $p_i$ is the predicted probability of leaf item $i$, and $Benefit(k)$ is the benefit of starting at item $k$.
For our menu-based interface, given a target item $l$, benefit of adaptively selecting a starting item $k$ is given by $Benefit(k) = p_k \cdot T(0,k)$.

With this formulation, we can predict the utility of every possible adaptation in the interface.
As this utility estimates the average selection time, we can then select the adaptation $a$ that minimizes this value:
\begin{equation}
    A = \underset{i \in n}{argmin}\ {U_i}
\end{equation}

\section{Walkthrough}


\subsection{Setup}

\emph{Menu Interface:}
We implement an application that allows users to select from a range of actions: watching TV shows or movies and listening to radio stations or specific music genres.
These actions are presented in a hierarchical menu, where a (sub-)menu and its contents are displayed in a list.
Additionally, a $\blacktriangleleft$ ``back'' button allows for corrective actions (\autoref{fig:menu_example}-a).
\autoref{fig:menu_example}-b illustrates the hierarchical structure of the menu interface and the sequence of actions taken by a user to navigate from a starting item to select a target item.

\emph{Probability Distribution:}
The XR device observes the user's context to generate a probability distribution over the given action space.
It can, for instance, observe the user's current location, activity, and time-of-day to make inferences about what their intentions would be.
We consider the following exemplary probability distribution for leaf items to demonstrate various scenarios:\\
$p$ = \texttt{\{Reality: 0.073, Comedy: 0.024, Drama: 0.098, Top 50: 0.024, New: 0.024, Classics: 0.122, News: 0.11, Charts: 0.085, Retro: 0.122, Jazz: 0.073, Electronic: 0.22, Rock: 0.025\}}.

\definecolor{s1}{rgb}{1.0,0.4,0.3}
\definecolor{s2}{rgb}{0.85,0.53,0.0}
\definecolor{s3}{rgb}{0.05,0.4,0.05}

\begin{table*}[htb]
\centering
\resizebox{\textwidth}{!}{%
\begin{tabular}{lrrrr}
\hline
\multicolumn{1}{c}{\textbf{Item}} & \multicolumn{1}{c}{\textbf{Probability}} & \multicolumn{1}{l}{\textbf{Scenario 1: Utility (in ms)}} & \multicolumn{1}{l}{\textbf{Scenario 2: Utility (in ms)}} & \multicolumn{1}{l}{\textbf{Scenario 3: Utility (in ms)}} \\ \hline
\textcolor{s2}{Entertainment}                     & 1.000                                    & 8121.0                                                   & \textcolor{s2}{\textbf{2121.0}}                                                   & 7605.0                                                   \\
Watch                             & 0.365                                    & 6731.0                                                   & 3461.0                                                   & 8050.0                                                   \\
\textcolor{s3}{Listen}                            & 0.635                                    & 4984.0                                                   & 2254.0                                                   & \textcolor{s3}{\textbf{5525.0}}                                                   \\
Shows                             & 0.195                                    & 6883.5                                                   & 5663.5                                                   & 10427.5                                                  \\
Movies                            & 0.170                                     & 7067.0                                                   & 5747.0                                                   & 10495.0                                                  \\
Radio                             & 0.317                                    & 4808.9                                                   & 4076.9                                                   & 7442.5                                                   \\
Music                             & 0.318                                    & 4736.5                                                   & 4008.5                                                   & 7114.5                                                   \\
Reality                           & 0.073                                    & 7687.2                                                   & 8125.2                                                   & 12441.0                                                  \\
Comedy                            & 0.024                                    & 8226.3                                                   & 8370.3                                                   & 12931.5                                                  \\
Drama                             & 0.098                                    & 7370.5                                                   & 7958.5                                                   & 11982.5                                                  \\
Top 50                        & 0.024                                    & 8299.2                                                   & 8443.2                                                   & 12996.0                                                  \\
New                            & 0.024                                    & 8294.4                                                   & 8438.4                                                   & 12972.0                                                  \\
Classics                        & 0.122                                    & 7152.8                                                   & 7884.8                                                   & 11674.0                                                  \\
News                               & 0.110                                     & 5857.0                                                   & 6517.0                                                   & 9659.0                                                   \\
Charts                               & 0.085                                    & 6120.0                                                   & 6630.0                                                   & 9849.0                                                   \\
Retro                              & 0.122                                    & 5673.8                                                   & 6405.8                                                   & 9283.0                                                   \\
Jazz                             & 0.073                                    & 6228.8                                                   & 6666.8                                                   & 9865.0                                                   \\
\textcolor{s1}{Electronic}                            & 0.220                                     & \textcolor{s1}{\textbf{4523.7}}                                                   & 5843.7                                                   & 7954.5                                                   \\
Rock                           & 0.025                                    & 6761.2                                                   & 6911.2                                                   & 10367.0                                                  \\ \hline
\end{tabular}%
}
\caption{Model results showing the list of menu items, probabilities, and utility values for the three scenarios. The value associated with the optimal adaptation for each scenario are shown in bold (\texttt{Electronic} = 4523.7 ms for scenario 1, \texttt{Entertainment} = 2121.0 ms for scenario 2, \texttt{Listen} = 5525.0 ms for scenario 3.)}
\label{tab:scenario_costs}
\end{table*}

\subsection{Example Scenarios}

Here, we present a set of illustrative scenarios that imitate different conditions or properties of the XR system, to study how they might result in varying model predictions and adaptations.  
For each scenario, \autoref{tab:scenario_costs} presents the model outputs (utility) for all menu items.


\subsubsection{Scenario 1: low inspection cost, high selection cost, low correction cost}


Here, the XR system is designed such that inspections have a low cost, corrections have a low cost, and selections have high cost.
This might be the case when the UI displays texts in large fonts, simple gestures are used for correction (e.g. swipe), and selection is tedious (e.g. point-and-click).
We use the following values: $T_{inspect}$ = 100 ms, $T_{select}$ = 2500 ms, $T_{correct}$ = 500 ms.

The system selects \texttt{Electronic} as the optimal adaptation.
Since this item has high probability and selection is more time-consuming that correction or inspection, it is advantageous for the system to make this \emph{optimistic} adaptation.

\subsubsection{Scenario 2: low inspection cost, low selection cost, high correction cost}

In this scenario, while inspection is similar to scenario 1, selection now has a low cost (e.g. pinch), while correction has a higher cost (e.g. point-and-click).
We use the following values: $T_{inspect}$ = 100 ms, $T_{select}$ = 500 ms, $T_{correct}$ = 2500 ms.

The system selects \texttt{Entertainment} as the adaptation to optimize the utility.
Given the current context,  with uncertainty in the contextual predictions, the system chooses the root item for a \emph{conservative} adaptation to avoid the high correction costs.

\subsubsection{Scenario 3: high inspection cost, average selection cost, average correction cost}

Finally, we consider a case where inspection costs are relatively high; this might be the case when text is rendered in small fonts, or items are represented using abstract icons.
Additionally, we select average selection and correction costs, where similar gestures are used for both actions (e.g. left-swipe and right-swipe).

Here, the system selects \texttt{Listen} as the adaptation.
By selecting a high-probability sub-menu, which is neither the root nor a leaf item, it considers the trade-offs to optimize the average  selection time using a \emph{balanced} adaptation.

\subsection{Summary}
As the above scenarios illustrate, our simulation model successfully takes various interaction costs into account to compute utility.
By doing so, it can select optimal adaptations, which vary between scenarios, that minimize the average selection time.
In contrast, to the above results, a greedy approach would always select the highest probability item (\texttt{Electronic}) as the adaptation.
\section {Discussion}

In this position paper, we have argued for a computational approach to adapt XR interfaces that considers both costs and benefits of adaptations. 
We presented a novel model that considered three cost components -- inspection, selection, and correction -- towards simulating user interactions with a hierarchical menu.
Given a contextual probability distribution, the model can estimate utility and select optimal adaptations that minimize interaction time.
Our illustrative scenarios demonstrate the benefits of such a model: by considering varying cost parameters, it can enable a system to select varying adaptations that are suitable for particular scenarios.

We see several exciting topics for future research on model-based adaptive interfaces.
While we consider a serial search strategy, an extended model could account for varying visual search strategies such as foraging and recall.
Additionally, model predictions can be improved by identifying user expertise from prior interactions and including it in the model.
Our model considers total interaction time as the objective for optimization; exploration of other aspects, such as ergonomics \cite{10.1145/3411764.3445349} and learnability \cite{doi:10.1177/154193128202600613} and their associated cost parameters will be important for optimal adaptive XR systems. 
Furthermore, extending the model to consider different adaptation styles will increase its applicability in various systems.
For instance, our model considers a hierarchical menu where a single item is recommended as the starting point.
Future adaptive systems can benefit by similarly developing simulation-based models for scenarios where multiple recommendations, or custom shortcuts, are presented.

To conclude, computational methods for generating or adapting their interfaces are a promising approach for the future of highly usable XR systems.
By adapting based on both contextual inference and interaction utility, they will be able to overcome new challenges imposed by the device and the environment.
We believe future applications can benefit from applying simulation-based interaction models, and look forward to developing and discussing novel models and approaches for such adaptive interfaces.

\bibliographystyle{ACM-Reference-Format}
\bibliography{References}


\end{document}